\documentstyle[graphicx,twocolumn,prl,aps]{revtex}

\def\etal{{\it et al.}}
\def\mn{m_{\scriptscriptstyle N}}
\def\HBCPT{HB$\chi$PT}

\begin{document}

\pagestyle{empty}
\narrowtext
\parskip=0cm
\noindent

{\bf Comment on ``Complete One-Loop Analysis of the Nucleon's Spin 
Polarisabilities"}

\smallskip

In a recent letter, \cite{gellas} Gellas \etal\ published results
for the spin polarisabilities of the nucleon calculated to
next-to-leading order (NLO) 
in heavy
baryon chiral perturbation theory (\HBCPT).  
Their results differ 
from those in Refs.~\cite{osbornetc}. Although this is due to
different definitions and not to discrepancies in computing Feynman diagrams,
it has caused some confusion in the community.  In this comment we show 
that it is the definition of Ref.~\cite{osbornetc}, and not that of Gellas 
\etal, which should be compared with the values extracted from dispersion 
relations (DR's) \cite{DR}.

The disagreement stems from the standard procedure of excluding 
the nucleon pole graphs from the definition of nucleon polarisabilities.  
In particular,
in fixed-$t$ dispersion relations each of the six independent 
Compton amplitudes is split
into two parts, a pole piece and a contribution from an integral over 
the imaginary part of the amplitude above the $\pi N$ cut; the latter is
obtained from photoproduction data.  The polarisabilities $\gamma_i$
are then defined as the lowest 
terms in the energy expansion of the second (non-pole) piece of the real part
of the amplitude.  The pole occurs when the
intermediate nucleon is on shell. The full pole contribution
can be calculated using Dirac spinors with the on-shell $\gamma N$ coupling,
and depends only on the nucleon charge, mass and magnetic moment.  There are no
``off-shell" or ``sideways" form-factors involved, and so it is gauge-invariant, 
Lorentz-covariant, and satisfies the DR's.
The low energy theorems (LETs) for Compton scattering are all pole terms.

\begin{figure}[t]
\centering
\includegraphics[width=0.18\textwidth]{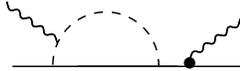}
\caption{ The disputed diagram.}
\end{figure}

The difference between the results 
of Refs. \cite{gellas,osbornetc} is in the 
treatment of the contribution from Fig.~1.   
If this is expanded in powers of the photon energy, $\omega$, the leading term
is needed to renormalise the bare value of the anomalous magnetic moment in the 
leading order (LO) pole terms, as might be expected.  (Less obviously, many
other  graphs also have leading terms which contribute to this correction.) 
The
disagreement between Gellas \etal\ and us is in the treatment of the 
$\omega^3$ terms from Fig.~1.  We include them in the
polarisabilities, they do not.

It is clear, however, that as the pole terms in the DR's depend only on
the on-shell couplings, any part of the contribution from Fig.~1 
(or other loop diagrams)
which is not purely a renormalisation of the mass or magnetic moment
is not part of the pole term. Thus it should be included in the definition of
the polarisabilities if comparison is to be made with the values extracted 
from DR's. Indeed by Cauchy's theorem the polarisability contributions from
the amplitude of Fig.~1 are related to an integral over the its imaginary part,
which arises from intermediate on-shell $\pi N$ states---just as in 
the similar graph where the second photon couples to the nucleon
inside rather than outside the pion loop.

For forward scattering, the polarisability $\gamma_0$, can be
directly related to the spin-dependent total inelastic
cross section via a GDH-style DR.  Clearly all diagrams contribute
to this. However as the pole contribution beyond the LET vanishes for forward
(and backward) angles, $\gamma_0$ will be the same whether defined this way
or via fixed-$t$ DR's.

We stress that this is not a problem of relativistic versus heavy-baryon 
formulations.  If however, like Gellas \etal, one wishes to propose a new
definition of polarisabilities excluding one-particle reducible 
graphs---polarisabilities which could not then, as we have argued, be directly
compared with values extracted from DR's---then care must be taken that the
definition is Lorentz and gauge invariant.
In  \HBCPT\, the procedure of Gellas \etal\ violates Lorentz
invariance to the order in the expansion in powers of $1/\mn$ to which they 
are working \cite{mcg}. Its gauge invariance has also not been demonstrated. 

It may be worth stressing here that while
\HBCPT\ will not give the pole at any finite order, 
it can reproduce the expansion of the amplitude
in powers of $1/\mn$. The true pole does not occur at physically
realisable values of the energy and momentum transfer, nor is it 
so close as to make the expansion badly behaved.  Thus \HBCPT\ 
can calculate the physical amplitude in the vicinity of zero energy 
without any problems of principle.  (Issues of chiral  convergence 
or the effects of the delta are irrelevant to this discussion, though 
they are important in practice.)

It is true that there are significant discrepancies between the polarizabilities 
from our NLO \HBCPT\ calculations and the DR analyses \cite{DR}.
Since the NLO corrections are of the same magnitude as the LO values 
however, there is no reason to expect that the series has converged.

\smallskip

\noindent 
Michael C. Birse$^*$, Xiangdong Ji$^\dagger$ and Judith A. McGovern$^*$\\

\vspace{-0.6cm}

\begin{quote}
{\small 
${}^*$Department of Physics and Astronomy,
University of Manchester, M13 9PL, U.K.\\
${}^\dagger$Department of Physics, University of Maryland,
College Park, MD20742
}\end{quote}



\vspace{-0.4cm}

\begin{references}

\vspace{-1.6truecm}

\bibitem{gellas} G. C. Gellas, T. R. Hemmert and U.-G. Mei\ss ner, 
Phys.\ Rev.\ Lett.\ {\bf 85} 14 (2000).

\bibitem{osbornetc} X. Ji, C-W.\ Kao and J. Osborne, Phys.\ Rev.\ {\bf D 61} 
074003 (2000);\\
K. B. V. Kumar, J. A. McGovern and M. C. Birse, 
{\tt hep-ph/9909442}; Phys.\ Lett.\ {\bf B 479} 167 (2000).

\bibitem{mcg} J. A. McGovern, M. C. Birse and K. B. V. Kumar, 
{\tt  nucl-th/0007015}

\bibitem{DR} 
D. Babusci \etal, Phys.\ Rev.\ {\bf C58} 1013 (1998);\\
D. Drechsel \etal, Phys.\ Lett.\ {\bf B 420} 248 (1998);\\ 
D. Drechsel \etal, Phys.\ Rev.\ {\bf C61} 015204 (2000).

\end{references}
\end{document}